%Paper: hep-th/9402146
%From: farina@vms1.nce.ufrj.br
%Date: Fri, 25 Feb 1994 14:56:33 -0300

\def\({\c c}
\def\|{\'\i }
\def\sqr#1#2{{\vcenter{\hrule height.#2pt
     \hbox{\vrule width.#2pt height#1pt \kern#1pt
      \vrule width.#2pt} \hrule height.#2pt}}}

\def\shpartial{{\slash\!\!\!\partial}}
\def\shD{{\slash\!\!\!\!D}}
\def\shA{{\slash\!\!\!\!A}}

\def\ni{{\noindent}}
\font\titulo = cmb10 scaled\magstep3

\baselineskip=18pt
\magnification=1200
\centerline{\titulo Chiral Bosons as solutions }
\vskip 0.5 true cm
\centerline{\titulo of the BV master equation }
\vskip 0.5 true cm
\centerline{\titulo for 2D chiral gauge theories}
\vskip 2 true cm
\centerline{\bf N. R. F. Braga and H. Montani$^\ast$}
\vskip 1 true cm
\centerline{Instituto de F\|sica}
\centerline{Universidade Federal do Rio de Janeiro}
\centerline{Rio de Janeiro  21945  Caixa Postal 68.528}
\centerline{Brasil}
\vskip 3 true cm
\centerline{\sl Abstract}
\vskip 0.5 true cm
\ni {\sl We construct the chiral Wess-Zumino term as a solution for the
Batalin-Vilkovisky master equation for anomalous
two-dimensional gauge theories, working in an extended
field-antifield space, where the gauge group elements
are introduced  as additional degrees of freedom.
 We analyze the Abelian and the non-Abelian cases, calculating in
both cases the BRST generator in order to show the physical
equivalence between this chiral solution for the master equation and the
usual (non-chiral) one.}

\vskip 1 true cm
\ni $\ast$ Present address: Centro At\'omico Bariloche; 8400 - S. C. de
Bariloche, Argentina.
\vskip 0.5cm
\ni PACS: 11.30.R, 11.15, 11.10.E
\vfill\eject

%\noindent{\bf\underbar {1-Introduction}}
%\vskip 1cm

The Batalin-Vilkovisky$^{[1]}$ scheme is a powerful
method to build up a BRST invariant
generating functional for a wide range of gauge theories. This scheme, also
called  field-antifield formalism, works on a configuration space that
includes, besides the original classical fields, a set of partners, called
antifields, sharing the opposite statistics. The introduction of these
antifields makes it possible to define a new operation, the
anti-bracketing, in such a way that
any BRST variation can be expressed in terms of the antibracket with an
extended action $W = S +  \sum_{j=1}^\infty\hbar^j M_j$ .
Where $S$ is the  classical part of the action and the $M_i$ are the
quantum corrections.
This action is obtained as the solution of the so called master
equation, which at the classical level ( zero order in $\hbar$ )
requires the vanishing of the antibracket of $S$ with itself.  The
higher order terms of the master equation are related, at least
formally, to the purely  quantum effects reflecting the non gauge
invariance of the path integral measure. In other words, quantum
correction to the master equation take into account the anomalies.
Moreover, the master equation is a systematic way to write down the
anomalous BRST Ward identities.  In fact, in a recent work, Troost,
van Nieuwenhuizen and Van Proeyen$^{[2]}$ have deeply analyzed the
problem of quantization of anomalous gauge theories in the
field-antifield framework, showing that the presence of a genuine
anomaly places an obstruction to the construction of a local solution
for the master equation at the quantum level. This was observed in
the quantization of the chiral Schwinger model, where we have
explicitly constructed a non local solution for the first quantum
contribution to the master equation, which turns local by the
introduction of an auxiliary field$^{[3]}$.

\ni Following the proposal of Faddeev and Shatashvili$^{[4]}$, and the
mechanism to generate the Wess-Zumino term$^{[5]}$ developed by
Babelon, Schaposnik and Viallet$^{[6]}$, and by Harada and
Tsutsui$^{[7]}$, we have recently proposed a mechanism to get rid the
obstruction in solving the master equation at the quantum level$^{[8]}$.
By considering that  the quantum fluctuation may render the gauge group
elements as dynamical variables, we extended the initial field-antifield
configuration space also including them. The new extended action $S$,
constructed  on this extended field-antifield space, must also
include in its Hessian matrix the generator of transformations of
the gauge group elements. So, the action $S$ contains an additional term
involving the corresponding antifield.
These term   allows one to overcome the topological obstruction,
giving rise to a local solution for the quantum master equation. By
analyzing what happens from the canonical point of view, we have
shown that the new term comes to restore the quantum nilpotency of the BRST
generator, and to eliminate the Schwinger term of the Gauss law
algebra$^{[8,9]}$. Also, it was shown that the action of the BRST operator
on the physical states annihilate a chiral sector of the bosonic
field associated to the gauge group element.
These idea was generalized  by Gomis and Paris in reference [10],
where by using the quasigroup structure of the gauge
generators$^{[11]}$, they find the general form of the transformation
of the group elements only requiring that the initial quasigroup structure
be preserved after the introduction of the new fields. They also
build up the general form of the solution of the master equation.

In this letter, we analyze  the master equation at the quantum level, for
Abelian and non Abelian gauge field theories, in the extended
field-antifield space, showing that it is possible to construct
 chiral Wess-Zumino terms as solutions for the master equation.
This result is similar to ref. [12], where it was
shown that the bosonised version of a chiral fermionic theory can be written
in terms of a chiral scalar particle.

\ni Let us start with the Abelian case, the Chiral Schwinger Model (CSM),
which is described by the action:

$$S_o = \int d^2x \{i \overline\psi\; \shD\;{(1-\gamma_5)\over 2}\;\psi
- {1\over 4} F_{\mu\nu}F^{\mu\nu} \}\eqno(1)$$

\ni The standard application of the Batalin-Vilkovisky  formalism leads to
a classical action  $S$ including terms involving the antifields and
the generators of the gauge symmetry:

$$S = S_o  + \int d^2x \{A^\ast_\mu \partial^\mu c
+ i\psi^\ast\psi c - i\overline\psi \;\overline{\psi^\ast} c \}
\eqno(2)$$

\ni This action satisfies the master equation at zero order in $\hbar$:
$(S,S) = 0 $, where the antibracket is defined as
 $(X,Y) = {\partial_rX\over
\partial\phi} {\partial_lY\over\partial\phi^\ast}
- {\partial_rX\over \partial\phi^\ast}
  {\partial_lY\over \partial\phi}$

\ni The master equation to first order in $\hbar$ reads

$$(M_1,S) = i \Delta S ,\eqno(3)$$

\ni where  $\Delta \equiv
{\partial_r\over\partial\phi^A}{\partial_l\over\partial\phi^\ast_A}\;$
the calculation of $\Delta S$, that  represents the BRST anomaly, is
explained in detail in references [3] and [8], resulting in:

$$\Delta S = {i\over 4\pi}\int d^2x\,\; c\; [ (1-a)\partial_\mu A^\mu -
\epsilon^{\mu\nu}\partial_\mu A_\nu ]\eqno(4)$$

\ni As was explained in references [2] and [3],
the master equation (3) does not
admit local solutions so, in this context,  it is not possible to
construct a gauge independent generating functional for the CSM.
Following the same reasoning of references [8],[9], and the
fact that a local effective action for
the CSM will include a dynamical extra field$^{[13]}$,
we enlarge the configuration space, adding the field $\theta$
and its  corresponding
antifield $\theta^\ast$. So, we get the extended classical action:

$$ S^\prime = S_o  + \int d^2x \{ A^\ast_\mu \partial^\mu c
+ i\psi^\ast\psi c - i\overline\psi \;\overline{\psi^\ast} c +
\theta^\ast \, c\}
\eqno(5)$$

\ni where we have considered that the gauge group parameter $\theta$
transforms as

$$ \theta\rightarrow\theta + \lambda \eqno(6)$$

\ni Observe that now the action $S^\prime$ depends on $\theta^\ast$
through the inclusion of the term $\theta ^\ast c$, and this extra term play a
fundamental role modifying the master equation (3). Now, one is able
to construct local solutions for it, depending also on the field $\theta$.
The new master equation at $\hbar$ order is

$$(M_1, S^\prime) =
\int \lbrace {\partial_r M_1\over\partial A^\mu}
{\partial_l S^\prime \over\partial A^\ast_\mu} +
 {\partial_r M_1\over\partial \theta}
{\partial_l S^\prime \over\partial \theta^\ast} \rbrace
= i \Delta S\eqno(7)$$

\ni It is worth remarking that the inclusion of the
$\theta$ field does not modify $\Delta S$, which is still given by
eq. (4).

It was shown in reference [9] that the usual Wess-Zumino
term for the CSM:

$$\overline M_1 = -{1\over 4\pi} \int d^2x \left\lbrace {(a-1)\over 2}\,
\partial_\mu \theta \; \partial^\mu \theta + \theta \left\lbrack (a-1)
\partial_\mu A^\mu + \epsilon^{\mu\nu} \partial_\mu A_\nu \right\rbrack
\right\rbrace\eqno(8)$$

\ni is a solution to this equation.  The canonical analysis of the theory
described now by the action $S^\prime + \hbar \overline M_1$, leads to the
modified Gauss law:

$$\overline\Omega = \Pi_1^\prime + J_0 - {\hbar\over 4\pi} [(a-1) A_0 + A_1]
- \Pi_\theta + {\hbar\over 4\pi} \theta^\prime\eqno(9)$$

\ni where $J_0$ is the chiral current defined by:

$$J_0 = \overline\psi (x)\gamma_0 {(1-\gamma_5)\over 2} \psi(x)$$

\ni The BRST operator can be obtained from the Gauss law (9) by the
relation $Q = \Omega c$, and its nilpotency may be verified from the
calculation of the anticonmutator $[ Q , Q ]_{_+}$, by taking into account
the non-trivial commutator$^{[14]}$:

$$i [\,\Pi_1^\prime(x), J_0(y)]_{_-} = {\hbar^2\over 4\pi}\;
\delta^\prime (x_1 - y_1)\eqno(10)$$

\ni However, the solution (8) is not unique. Another interesting
solution for the master equation (7) is:

$$\eqalignno{ M_1 = {1\over 4\pi} \int d^2x \{
&- \theta^\prime \dot\theta + (a-1)(\theta^\prime )^2 +
2 \theta^\prime [ A_o - (a-1) A_1 ]\cr  &- {1\over 2} (a-1) A_o^2
- A_1A_o + {3\over 2} (a-1) A_1^2 \}
 &(11)\cr }$$

\ni that represents the action of a chiral boson coupled to the gauge field.
The are no higher order contributions to the master equation, so
the quantum action is just

$$W = S^\prime + \hbar M_1$$

\ni This Wess-Zumino term is the same as the action obtained in reference
[12], after the so called chiral bosonization of the chiral Schwinger model.
However, it should be emphasized that, in this reference, the chiral
constraint is imposed to the functional generator by means of a
delta functional  in momentum space based on the reasoning that only
one chirality of a scalar field should  be necessary in order to
represent a chiral fermion.  In the present case, we obtain a
Wess-Zumino term that involves a chiral scalar field, just choosing
one of the solutions of the Batalin-Vilkovisky master equation.

\ni Let us now investigate the physical space of states, by analyzing
the consequences of this $M_1$, from the canonical point of view .
Calculating the canonical momenta associated to action (11), we get the
 chiral constraint  $\Pi_\theta =
-{\hbar \over 4\pi} \theta^\prime$ and the other primary constraint:

$$ \Pi_0 = 0  \eqno(12)$$

\ni and, building up the Hamiltonian, the time evolution
of (12) leads to the secondary constraint:

$$ \Omega = \Pi_1^\prime + J_0 - {\hbar \over 4\pi}[(a-1)
A_o - A_1 ] + {\hbar \over 2\pi}\theta^\prime\eqno(13)$$

\ni that, by using the $\Pi_\theta$ definition, it acquires the same form
as (9)

$$ \Omega = \Pi_1^\prime + J_0 - {\hbar \over 4\pi}[(a-1)
A_o - A_1 ] - (\Pi_\theta - {\hbar \over 4\pi}\theta^\prime)$$

\ni The canonical generator of the BRST transformations is thus:

$$ Q = \int dx_{_1} \;\Omega (x) \,c(x) \eqno(14)$$

\ni In order to investigate the quantum nilpotency of Q, we will
build up the anticommutator $[ Q , Q ]_+ = 2Q^2$.

$$\eqalign{\lbrack Q , Q \rbrack_{_+}\, =  \int dx_1 \int
dy_1 \; \lbrace c(x)\,&
[ \pi^\prime_1(x) + J_0(x) + {\hbar \over 4\pi} A_1(x) ,
\pi^\prime_1(y) + J_0(y) + {\hbar \over 4\pi} A_1(y) ]_{_-} \cr
&+ [ \pi_\phi(x) - {\hbar \over 4\pi}\phi^\prime(x) ,
\pi^\prime_1(y) - {\hbar \over 4\pi}\phi^\prime(y) ]_{_-}\,
 c(y)\rbrace \vert_{_{x_o = y_o}}\cr}\eqno(15)$$

\ni This anti-commutator depends on the equal time commutator
of the chiral current that requires a
careful management at the quantum level as it involves the
product of operators at the same point .
A regularization scheme is then needed in order to  overcome this problem.
As it is well known, the calculation of this
commutator generates the Schwinger term, transforming the current
algebra in a Kac-Moody one.
By using the Bjorken, Jonhson, Low limit$^{[15]}$,
one is able to relate the
vacuum expectation value of this current commutator to the second functional
derivative of the effective action $S_{_{eff}}$,
which is equivalent to the logarithm of the determinant of
the fermionic operator
$[\shpartial + \shA {(1-\gamma_5)\over 2}]$ $^{[16-19]}$.
Thus, following ref. [19], one gets the standard result:

$$i[J_0(x) ,J_0(y) ]_{_-} =
{\hbar^2\over 2\pi}\;  \delta^\prime (x_1 - y_1) \eqno(16) $$

\ni Furthermore, the non-trivial Jacobian of the
transformation generated by $J_0$ gives rise to a Schwinger term  also in
the commutator between $J_0$ and $\Pi_1^\prime$ [eq. (10)], as shown in ref.
[14].

\ni The appearance of these Schwinger terms would means the
breakdown of the gauge symmetry and would  also destroy the nilpotency of
the operator $Q$ .  However, the presence of the extra field $\theta$ leads
to the appearance of the two other commutators in (14).  These commutators
are trivial, in the sense that they are not anomalous. Introducing these
results in (14) one gets

$$[ Q , Q ]_{_+} = 0 $$

\ni showing explicitly that the introduction of the field $\theta$
leads to a theory that has a BRST generator that is nilpotent at the
quantum level.

\ni It is important to remark that, comparing the present results with
those from reference [9], we see that the BRST generator is essentially
the same in the chiral and the non-chiral solution for the master equation
(7). As already pointed out in [9] $Q$ involves the chiral constraint
$\dot\theta = \theta^\prime$. Therefore, the physical states, that are
annihilated by $Q$ involve just chiral scalars, in any formulation.

Now we will consider the non-Abelian gauge field theory in two dimensions,
that means: $QCD_2$. The classical action, including gauge fixing terms,
is$^{[8]}$:

$$ S = \int d^2x \lbrace i\overline\psi\;\shD {(1-\gamma_5)\over 2}\,\psi
- {1\over4} Tr(F^{\mu\nu}
F_{\mu\nu} \rbrace +\; \lbrace A^\ast_\mu [D^\mu,c] +
\psi^\ast c \psi + \overline\psi c
\overline\psi^\ast  + c^\ast\wedge c\wedge c\rbrace\eqno(17)$$

\ni This equation satisfies the master equation at order zero
in  $\hbar$ : $(S,S) = 0 $. Now, to first order in  $\hbar$
the master equation involves $\Delta S$.
One can see that $\Delta$ affects all the terms in
$S_o$ and a naive calculus yields $\Delta S\propto c\,\delta (0)$.
The arising of this ill-defined expression corresponds to the well-known
fact that one needs a regularization to define the measure$^{[1,2]}$.
This calculation, already explained in references [3,8], yields

$$\Delta S = {i\over 4\pi} \int d^2x\; tr\lbrace c\, \lbrack
\varepsilon^{\mu\nu} \partial_\mu A_\nu -
\partial_\mu A^\mu\rbrack\rbrace\eqno(18)$$

\noindent where one can identify the consistent gauge anomaly.
The master equation to $\hbar$ order is
essentially the anomalous BRST Ward
identity and, as it is well known, it does not admit local solutions.

\ni As already explained, in order to build up local solutions to the
master equation, we introduce the extra field h, associated to the
gauge group. Following reference [8] we add to the classical action
(17) the gauge fixing term associated with the new field, so that

$$S^\prime = S + \int dx^2 \, h^\ast h c $$

\ni Thus, the path integral measure also must include the measure
$Dh$, which stands for the local product of Haar measures on G.
The higher
order contributions to the master equation take account of the variation
of the path integral measure under the above symmetries, through the
evaluation of $\Delta S^\prime$.
Observe that, because of the trivial gauge invariance of
the Haar measure, the transformation $s_o h = hT^ac_a$ leaves the
measure $Dh$ invariant, which means that the corresponding Jacobian
admits a covariant regularization rendering it a trivial one.
So, the only non-trivial Jacobian come again from the fermionic measure.
Then, the $\Delta S$ is still given by the expression (18).

\ni In reference [8] we found a solution to
$(M_1,S^\prime) = i \Delta S^\prime$ of the  form

$$\eqalign{\overline M_1(A_\mu,h) = \Gamma [h] + \int d^2x\; tr \lbrace
&{1\over 2}(\alpha + {1\over 4\pi})\; \partial_\mu h^{-1} \partial^\mu h +
\alpha\, h^{-1}\partial_\mu h\; A^\mu\cr
& - {1\over 4\pi}\, \varepsilon^{\mu\nu}
\,h^{-1} \partial_\mu h\; A_\nu - {1\over 2} (\alpha +
{1\over 4\pi})\, A_\mu A^\mu \rbrace}\eqno(19)$$

\noindent where $\alpha$ is an arbitrary parameter and $\Gamma [h]$ is
the Wess-Zumino-Witten action$^{[4,9]}$:

$$\Gamma [h] = {-1\over 8\pi} \int d^2x\;
tr \,(\partial_\mu h^{-1} \partial^\mu h)\;
+ {1\over 4\pi} \int^1_0 dr \int d^2x\; \varepsilon^{\mu\nu}\,
tr ( h^{-1}_r \partial_r h_r\; h^{-1}_r \partial_\mu h_r
\;h^{-1}_r \partial_\nu h_r )\eqno(21)$$

\noindent with $h_r$ being some interpolation between $h_0 = 1$
and $h_1 = h$
 This solution  corresponds to the usual Wess-Zumino
term for the chiral $QCD_2$.
As in the Abelian case, it is possible to build up a chiral solution
for the master equation, that corresponds to a non abelian chiral boson
coupled to the gauge field$^{[20]}$:

$$\eqalign{ M_1(A_\mu,h) =&  \int d^2x\; tr \lbrace
{1\over 4\pi}\;(h^{-1} \partial_0 h h^{-1} \partial_1 h +
 h^{-1}\partial_1 h h^{-1} \partial_1 h) \;\cr
& + {1\over 4\pi}\, ( -2h^{-1} \partial_1 h A_+ - {1\over 2}A_0^2
+ A_0 A_1 +{3\over 2}A_1^2 )\rbrace \cr
&+ {1\over 4\pi} \int^1_0 dr \int d^2x\; \varepsilon^{\mu\nu}\,
tr ( h^{-1}_r \partial_r h_r\; h^{-1}_r \partial_\mu h_r
\;h^{-1}_r \partial_\nu h_r )}\eqno(22)$$

\ni Again, it should be stressed that we did not impose the chiral
constraint. We are just choosing one of the possible solutions of the
master equation.
As in the abelian case, the higher order contributions to the master
equation can be taken as zero. So, the quantum action is just
$W = S^\prime + \hbar M_1$

\ni Now we analyze the action $W$ from the canonical point of view,
building up the BRST generator.
The main consequence of the additional term $M_1$ is to modify the
Gauss Law so that now it reads:

$$ \Omega^a = D_1^{ab}\Pi_1^b + J_0^a - {\hbar \over 8\pi}
[A_0^a - A_1^a ] - {\hbar \over 2\pi} tr(h^{-1} \partial_1 h\,
T^a) + f^{abc} \Pi_{gh}^b c_d \eqno(23)$$

\ni with $J_\mu^a$ being the chiral current

$$J_0^a = \overline\psi \gamma^\mu ({1-\gamma^5\over 2}) T^a \psi $$

\ni Then, following the standard recipe for constructing the  BRST generator
from the constraints, we get:

$$\eqalign{ Q &= Q_o + Q^\prime \cr
&= \int dx_1 \lbrace  J_0^a c_a - {1\over
 2} [D^\nu , F_{o \nu} c] + {1\over 2} A^{\ast a}_o f^{abc} c^b c^c  -
{\hbar \over 2\pi} tr(h^{-1}\partial_1 h\, c) - {\hbar \over 2\pi} (A_0^a -
A_1^a)\, c_a \rbrace\cr} \eqno(25) $$

\ni where $Q_o$ stands for the BRST generator arising from the action (17) and
$Q^\prime$ is the contribution coming from (22).
If the chiral constraint is explicitly implemented, the term involving
the field $h$ can be written exactly as the one in reference 8 where the BRST
charge associated to the non-chiral Wess-Zumino term was build up.

\ni In checking the nilpotency of $Q$, we need
the commutator of the chiral current for the non-Abelian case$^{[17]}$:

 $$[ J_0^a(x) , J_0^b(x^{\prime})]_{_-} =
 -f^{abc} J_0^c(x) \delta (x_1 - x^{\prime}_1)
+ {i\over 4 \pi} \delta^{ab} \delta^\prime (x_{_1} - x^{\prime}_{_1})
 + {i\over 8\pi} f^{abc} (A_o - A_1)^c \delta(x_{_1}-x_{_1}^\prime)\eqno(27)$$

\ni together with the non-trivial commutator

$$[\Pi_1^{a\prime}(x), J_0^b(y)] = {\hbar^2\over 4\pi}
\delta^{ab}\, \delta^\prime (x_1 - y_1)$$

\ni and the others commutator being the usual ones. In this way it is
immediate to verify that

$$ Q^2 = {1\over 2} [ Q , Q ]_{_+} = 0 \eqno(28)$$

%%%%%%%%%%%%%%%%%%%%%%%%%%%%%%%%%%%%%%%%%%%%%%%%%%%%%%%%%%%%%

In conclusion we have shown that when applying the Batalin-Vilkovisky
quantization procedure to chiral gauge theories,
one is able to choose, among the various possible solutions
of the master equation, a solution involving a chiral Wess-Zumino field.
Therefore chiral Wess-Zumino terms are generated without imposing
the chiral constraint by hand, as usually done in the literature.
We have also calculated the BRST generator for the Abelian and non-Abelian
cases. The nilpotency of this operator at the
quantum level was shown for both cases.
This ensures that the physical states can be defined by the cohomology class
of this operator.  It was also pointed out that these generators are the
same as in the case of the standard (non chiral) Wess-Zumino terms, leading
us to the conclusion that the spaces of physical states are the same.

\vskip 2cm

\noindent Acknowledgements: This work is partially supported by
CNPq - Brasil. H. Montani thanks to R. Trinchero for fruitfull discussions.

\vfill\eject

{\bf \underbar {References}}
\vskip 0.8 true cm

\item{[1]-} I. A. Batalin and G. A. Vilkovisky, Phys. Lett. B102
(1981) 27, Phys. Rev. D28 (1983) 2567.
\item{[2]-} W. Troost, P. van Nieuwenhuizen and
A. Van Proeyen, Nucl. Phys. B333 (1990) 727.
\item{[3]-} N. R. F. Braga and H. Montani, Phys. Lett. B264 (1991) 125.
\item{[4]-} L. D. Faddeev, Phys. Lett. B145
(1984) 81; L. D. Faddeev and S. L. Shatashvili, Phys. Lett. B167
(1986) 225.
\item{[5]-} J. Wess and B. Zumino, Phys. Lett. B37 (1971) 95
\item{[6]-} O. Babelon, F. A. Schaposnik and C. M. Viallet, Phys.
Lett. B177 (1986) 385.
\item{[7]-} K. Harada and I. Tsutsui, Phys. Lett. B183 (1987) 311.
\item{[8]-} N. R. F. Braga and H. Montani, Int. J. Mod. Phys. A8 (1993)
2569.
\item{[9]-} N. R. F. Braga and H. Montani,  Phys. Rev. D, to be published.
\item{[10]-} J. Gom\|s and J. Paris, Nucl. Phys. B 395 (1993) 288.
\item{[11]-} I. Batalin, J. Math. Phys. 22 (1981) 1837.
\item{[12]-} K. Harada, Phys. Rev. Lett. 64 (1990) 139.
\item{[13]-} R. Jackiw and R. Rajaraman, Phys. Rev. Lett. 54 (1985) 1219.
\item{[14]-} R. C. Trinchero, Mod. Phys. Lett. A4 (1989) 187; C. Fosco and
R. C. Trinchero, Phys. Rev. D41 (1990) 1216.
\item{[15]-} J. Bjorken, Phys. Rev. 148 (1966) 1467;\par
K. Johnson and F. Low , Prog. Theor. Phys. (Kyoto), Suppl. 37-38
(1966) 74.
\item{[16]-} R. Jackiw, ``Field Theoretic Investigations in Current
Algebra", in {\sl Lectures on Current Algebra and Its Applications},
S. Treiman, R. Jackiw and D. Gross eds., Princeton University Press,
Princeton, NJ (1972).
\item{[17]-} S. Jo, Nucl. Phys. B259 (1985) 616.
\item{[18]-} R. E. Gamboa Sarav\|, F. A. Schaposnik and
J. E.Solomin, Phys. Rev. D33 (1986) 3762.
\item{[19]-} M. V. Manias. M. C. von Reichenbach, F. A. Schaposnik
and M. Trobo, J. Math. Phys. 28 (1987) 1632.
54 (1985) 1219.
\item{[20]-} J. Soneschein, Nucl.Phys. B309 (1988) 752.
\bye